# Constructing Receiver Signal Points using Constrained Massive MIMO Arrays

Markus Staudacher*, Gerhard Kramer*, Wolfgang Zirwas†, Berthold Panzner†, Rakash Sivasiva Ganesan†

*Institute for Communications Engineering, TUM

†Nokia Bell Labs, Munich

*Abstract*—A low cost solution for constructing receiver signal points is investigated that combines a large number of constrained radio frequency (RF) frontends with a limited number of full RF chains. The constrained RF front ends have low cost and are limited to on/off switching of antenna elements and a small number of phases. Severe degradations are typically observed for multi-user MIMO for these simple on/off antenna arrays. A few full RF frontends are shown to compensate for the signal errors of the high number of constrained RF frontends for various scenarios. An algorithm for such a hybrid RF (HRF) system is developed that achieves performance close to that of exhaustive search with respect to the mean square error of the constructed receiver signals for Rayleigh fading and the WINNER 2 Urban Macro channel model.

## I. Introduction

Massive multiple-input-multiple-output (MIMO) promises to deliver high spectral efficiency with low energy consumption [1], [2]. The theory for massive MIMO is studied in [3], [4] and implementation issues such as hardware costs and dimension are assessed in [5]. An uplink system with 1 bit analog-to-digital converters (ADC)s at the receiver approaches capacity when using QPSK [6], [7], [8], [9]. Several approaches simplify the implementation of massive MIMO while preserving some of the gains. For example, a simple idea is to use only one antenna at the base station (BS) at a certain time instant to transmit [10] so that the transmitter (Tx) needs only one RF-chain. As another example, hybrid beamforming reduces the total number of RF-chains to decrease the cost and power consumption of the BS [11].

We propose a design where a large number of low cost constrained RF-chains (CRF)s cooperate with a small number of full RF-chains (FRF)s. In the simplest case the CRFs use simple on/off switching, thereby requiring minimum functionality like a single bit DAC, a power amplifier (PA) with relaxed linearity constraints to achieve high power-added efficiencies (PAE), less stringent filter requirements, etc. One interesting use case is to add booster arrays with a large number of CRFs to existing macro sites - the FRFs - to form a HRF massive MIMO array. An HRF system can compensate for all the non idealities of the CRFs, as long as the number of FRFs is larger than that of the served data streams.

Compared to hybrid beamforming one can avoid the analog network and retain full precoding flexibility, which is limited for wideband analog beams.

Here we investigate a system with low- or one-bit DACs together with or without an additional phase shifter per CRF. Our contributions are twofold.

- We present multi-user (MU) MIMO precoding with CRFs and provide an algorithm that achieves close to optimal results with respect to the mean square error (MSE).
- We develop an HRF system and evaluate its benefits like shared cost, reduced power consumption and effects on the number of overall antenna elements.

*Notation*: We use boldface lowercase and uppercase letters to denote column vectors and matrices, respectively.

## II. 1 bit Massive MIMO Transmitter

### A. System Model

Consider a one-cell downlink with $K$ single antenna users and one base station (BS) equipped with $M$ antennas with $M \gg K$. The discrete-time complex received signal is

$$\boldsymbol{y} = \boldsymbol{H}\boldsymbol{x} + \boldsymbol{n} \quad (1)$$

where $\boldsymbol{H} \in \mathbb{C}^{K \times M}$ is the channel matrix from the BS to the $K$ users. The entry $h_{ij}$ of $\boldsymbol{H}$ is the channel coefficient between the j-th antenna of the BS and the i-th user equipment (UE). We consider Rayleigh fading where the channel coefficients $h_{ij}$ are independent $\mathcal{CN}(0,1)$ random variables, and the WINNER 2 Urban Macro channel model [12], where we define the $h_{ij}$ with QuaDRiGa [13]. $\boldsymbol{x} \in \mathbb{C}^{M \times 1}$ is the transmit signal that is constrained as will be explained in the following section. The entries of $\boldsymbol{n} \in \mathbb{C}^{K \times 1}$ are independent circularly symmetric Gaussian random variables.

We propose to construct receive signal points by turning on or off selected antenna elements of a massive MIMO array, potentially in combination with adapting the Tx-signal phase information (PS) per antenna element. The desired RX-values for multiple UEs are generated via superposition of the Tx-signals affected by the antenna specific channel components. This is depicted in Figure 1. The upper RF-chain resembles the elements of a FRF and the lower one shows the proposed CRF. In both cases we have baseband processing (BBP). The FRF signal is transformed to the analog domain with a standard DAC and converted to RF with the local oscillator (LO) and fed into a power amplifier (PA). The analog bandpass filter before the Tx-antenna suppresses out-of-band emissions.

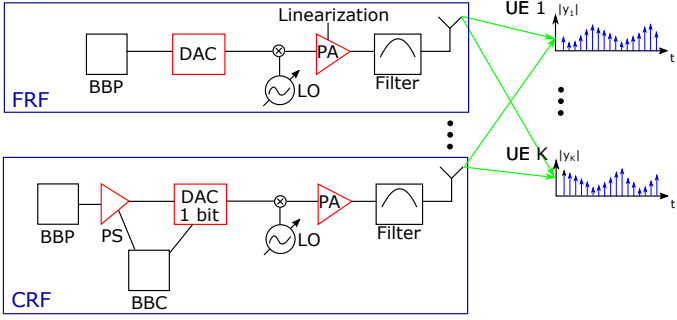

Fig. 1: FRF and CRF combination for signal generation at the receiver.

For the CRF we separately compute the phase and amplitudes in the baseband controller (BBC), i.e., whether the antenna should be turned on or off and what phase value should be used, depending on the desired signal values at the receivers. The BBC must take the limited number of phase shifter (PS) and amplitude values into account.

We compare two options where either CRFs generate the Rx-signals or an HRF system that uses both CRFs and FRFs. The time domain samples are assumed to be the time domain representation of OFDM signals. The evaluation criterion is the deviation - or minimum mean square error (MMSE) - between the intended and the realized time domain signals. Note that, so far, we assume perfect channel knowledge from all antenna elements to all UE Rx antennas.

We expect significant benefits with respect to cost, size and power consumption for CRFs compared to FRFs as the DAC might be omitted completely and the PA needs no linearization so that a much higher PAE can be achieved. By combining many active antenna elements, the power per antenna can be small, allowing for single chip or system-on-chip solutions. One may also reduce the requirements for otherwise bulky filter elements.

Figure 2 shows two possible antenna configurations. On the left side we have an HRF with CRFs that are supported by a few FRFs. On the right side an already existing FRF array is supported by a large array of CRFs to form an HRF array. The idea is to reuse available macro sites and to add *boosterarrays*, e.g., at the walls of the building.

### B. Transmission Scheme

We aim to minimize the MSE of UE receiver signals and for now ignore additive white Gaussian noise (AWGN) to evaluate upper performance bounds of the described schemes. The optimization problem can be expressed as

$$\text{minimize} \quad ||\boldsymbol{Hx} - \boldsymbol{u}||^2 \quad (2)$$
$$\text{subject to} \quad \boldsymbol{x} \in \left\{0, \frac{1}{\sqrt{M}}\right\}^M$$

where $\boldsymbol{u} \in \mathbb{C}^{K \times 1}$ contains the desired symbols of the UEs, with $E[|u_k|^2] = 1$. The entries of $\boldsymbol{x}$ are constrained to be either zero or $\frac{1}{\sqrt{M}}$, which ensures that the total

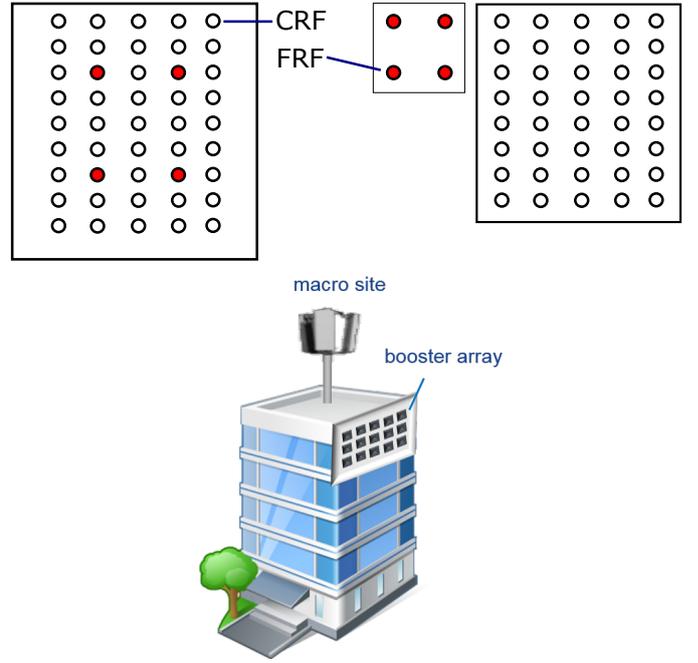

Fig. 2: Use cases for the CRFs.

**Algorithm 1** KS for $K \geq 1$, UEs no phase information

1: $\boldsymbol{H} = \boldsymbol{H}/\sqrt{M}$
2: $\boldsymbol{err} = \boldsymbol{u}$
3: **for** $i = 1 : M$ **do**
4:     $j^* = \text{argmin } ||\boldsymbol{err} - \boldsymbol{H}(:,j)||$
5:     **if** $||\boldsymbol{err}|| < ||\boldsymbol{err} - \boldsymbol{H}(:,j^*)||$ **then** stop;
6:     $\boldsymbol{err} = \boldsymbol{err} - \boldsymbol{H}(:,j^*)$
7:     $\boldsymbol{x}(j^*) = \frac{1}{\sqrt{M}}$
8:     $\boldsymbol{H}(:,j^*) = NaN$
9: **end**

power $P$ of the array is 1 when all antennas transmit. The constrained set can be extended if phase modulation is allowed, e.g., for 1 bit for the phase information the constraint is $\boldsymbol{x} \in \left\{0, \frac{1}{\sqrt{M}}, -\frac{1}{\sqrt{M}}\right\}^M$.

The problem (2) is known as binary least squares [14] and is related to the knapsack (KS) problem [15]. The problem is non-deterministic polynomial-time (NP) hard [15] and can (of course) be solved by using exhaustive search (ES), i.e., computing the MSE for every possible vector $\boldsymbol{x}$ and choosing the best one. As this is computationally very demanding, we resort to a suboptimal solution. Instead of finding the solution in one step, the algorithm iteratively activates the antenna that minimizes the remaining error at the receiver.

The algorithm is initiated by multiplying $\boldsymbol{H}$ with $\sqrt{P}$ and initializing the error with the desired receive vector $\boldsymbol{u}$, as all antennas are turned off at the beginning. In the next step the algorithm calculates the index of the column with minimal

**Algorithm 2** KS for $K \geq 1$, UEs with phase information
1: $\boldsymbol{H} = \boldsymbol{H}/\sqrt{M}$
2: $\boldsymbol{H} = \boldsymbol{v}_{ph}^T \otimes \boldsymbol{H}$
3: $\boldsymbol{err} = \boldsymbol{u}$
4: **for** $i = 1 : M$ **do**
5: $\quad j^* = \text{argmin } \|\boldsymbol{err} - \boldsymbol{H}(:,j)\|$
6: $\quad$ **if** $\|\boldsymbol{err}\| < \|\boldsymbol{err} - \boldsymbol{H}(:,j^*)\|$ **then** stop;
7: $\quad \boldsymbol{err} = \boldsymbol{err} - \boldsymbol{H}(:,j^*)$
8: $\quad j_{mod} = \text{mod}(j^*, M)$
9: $\quad ph_{index} = \text{floor}((j^* - 1)/M) + 1$
10: $\quad$ **if** $j_{mod} == 0$ **then** $j_{mod} = M$
11: $\quad \boldsymbol{x}(j_{mod}) = \frac{1}{\sqrt{M}} \boldsymbol{v}_{ph}(ph_{index})$
12: $\quad$ **for** $a = 0 : 2^m - 1$ **do**
13: $\quad\quad \boldsymbol{H}(:, aM + j_{mod}) = NaN$
14: **end**

Euclidean distance to the error vector. If the updated error is larger than the new error, the algorithm stops. Otherwise the error vector is updated, the antenna corresponding to the column of $\boldsymbol{H}$ is activated and the column of $\boldsymbol{H}$ is set to "not a number" (NaN) in order to be ineligible in the following iterations. The algorithm has a complexity of $O(M^2 \log(M) + M^2 K)$.

The algorithm can be extended to include phase information by calculating $\boldsymbol{H} = \boldsymbol{v}_{ph}^T \otimes \boldsymbol{H}$, where $\boldsymbol{v}_{ph}$ is a row vector containing all possible phase options and $\otimes$ is the Kronecker product. In Algorithm 2, $m$ is the number of phase bits, mod is the Modulo operation and floor is the floor operation.

*C. HRF Scheme*

Using CRF-chains leads to either a transmit power saving and/or a reduction in the MSE. For the HRF scheme the residual MSE should be either cancelled or reduced. Our approach is as follows. We apply the knapsack algorithm to the CRF antennas to find a solution close to the desired Rx-signal vector and then use the remaining FRF antennas to minimize the remaining error. Formally this can be described as follows: Solve (2) with the knapsack algorithm by replacing $\boldsymbol{H}$ by $\boldsymbol{H}_1 \in \mathbb{C}^{K \times M_1}$, $M$ by $M_1$ and $\boldsymbol{x}$ by $\boldsymbol{x}_1 \in \mathbb{C}^{M_1 \times 1}$, where $M_1$ is the number of CRF-chains. Then calculate the zero forcing (ZF) solution if $M_2 \geq K$:

$$\boldsymbol{x}_2 = \boldsymbol{H}_2^H \left(\boldsymbol{H}_2 \boldsymbol{H}_2^H\right)^{-1} \boldsymbol{u}_t \quad (3)$$

where $\boldsymbol{x}_2$ is the transmit vector and $\boldsymbol{H}_2$ the matrix of channel coefficients belonging to the set of FRF-chains. The vector $\boldsymbol{u}_t = \boldsymbol{u} - \boldsymbol{H}_1 \boldsymbol{x}_1$ contains the remaining error.

When $M_2 < K$ we use the least squares solution:

$$\boldsymbol{x}_2 = \left(\boldsymbol{H}_2^H \boldsymbol{H}_2\right)^{-1} \boldsymbol{H}_2^H \boldsymbol{u}_t. \quad (4)$$

Note that we do not impose power constraints on the antennas connected to the FRF-chains, as we want to observe the reduction of power usage due to the added CRF antenna array.

TABLE I: General system settings for the simulations

| | |
|---|---|
| # BS | 1 |
| # BS antennas ($M$) | 1-120 |
| BS sum power constraint ($P$) | 1 |
| # UE ($K$) | 1-10 |
| # UE antennas | 1 |
| Input alphabet | 256 QAM |
| Quantization scheme | 1 bit amplitude, 1-6 bit phase |
| Channel model | Rayleigh, WINNER II UMa |

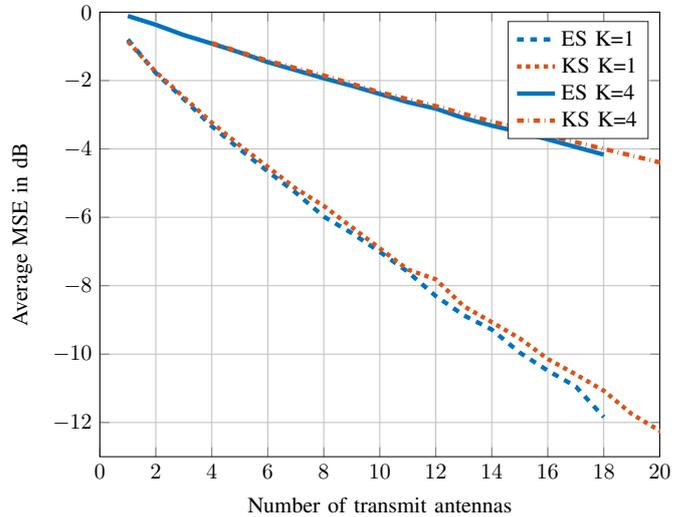

Fig. 3: Average MSE per UE for Rayleigh fading, $K = 1$ and for exhaustive search and the knapsack algorithm.

## III. SIMULATION RESULTS

We compare our schemes under Rayleigh fading and the WINNER 2 Urban Macro channel model for various scenarios. We use Monte Carlo simulations to compare the schemes in terms of MSE and transmit power.

*A. Modulation over the air*

The simulation parameters are given in Table I. To get to a sufficient symbol error rate for 256 QAM we target a MSE of -20 dB.

Figure 3 shows the performance of the ES and the KS algorithm. For complexity reasons the simulation is limited to 20 transmit antennas. For the simulated range the MSE difference is less than 1 dB, which seems remarkable considering the much lower complexity of the knapsack algorithm.

In Fig. 4 we depict the average MSE in dB for $K = 1$ UE and different numbers of phase bits $m$. The MSE is averaged over 1000 channel realizations and the different symbols of the 256 QAM alphabet. For just one UE the target of $-20$ dB can easily be achieved with about 40 transmit antennas and without phase. The convergence behavior is important. The MSE first declines exponentially fast. There is a point however where the available channel coefficients the algorithm can choose from have large absolute values as compared to the remaining error. Therefore we can see the convergence of the MSE at some point. Furthermore the addition of only 1 bit for phase information can improve the MSE by as much as 18 dB for 20 transmit antennas. Figure 5 shows the same scenario, but with $K = 10$ UEs. Clearly, the number of transmit antennas needed to achieve an MSE of $-20$ dB is much larger than before. Now with 1 bit phase information 105 transmit antennas are needed to fulfill the goal.

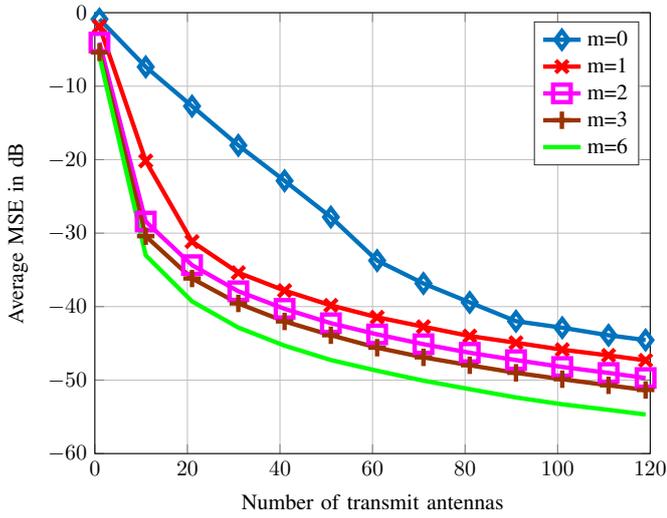

Fig. 4: Average MSE per UE for Rayleigh fading. $K = 1$ and $m$ bit for the phase.

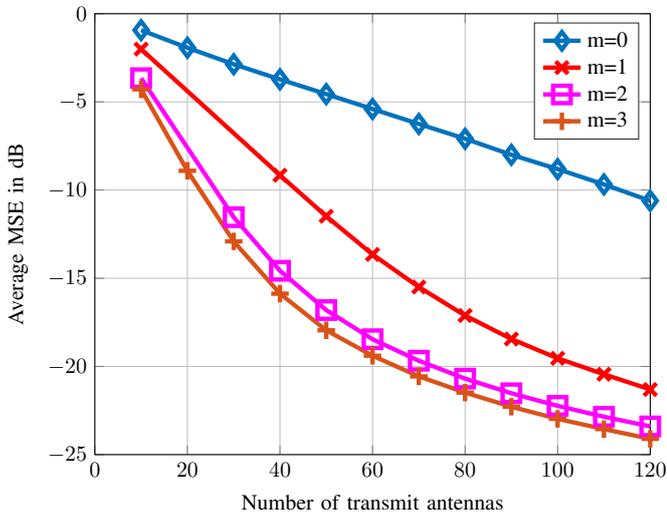

Fig. 5: Average MSE per UE for Rayleigh fading. $K = 10$ and $m$ bit for the phase.

### B. HRF Scheme

Consider the combination of many CRF-chains and some FRF-chains. We do not impose any power constrains on the FRF-chains, hence we can perfectly create the desired symbol at the receiver as long as $M_2 \geq K$. Figure 6 shows the transmit power of the FRF-chains for 3 different scenarios. In the first case we do not have enough RF-chains to construct the desired symbol perfectly. We start with a total of 20 transmit antennas of which 8 are connected to FRF-chains. As we increase the number of CRF antennas, the transmit power decreases from 3 dB at 20 antennas to -6 dB at 120 antennas. A similar behavior is observed for the system with 12 FRF-chains. For $K = M_2$ we observe a much higher transmit power as there are not enough degrees of freedom for the ZF solution.

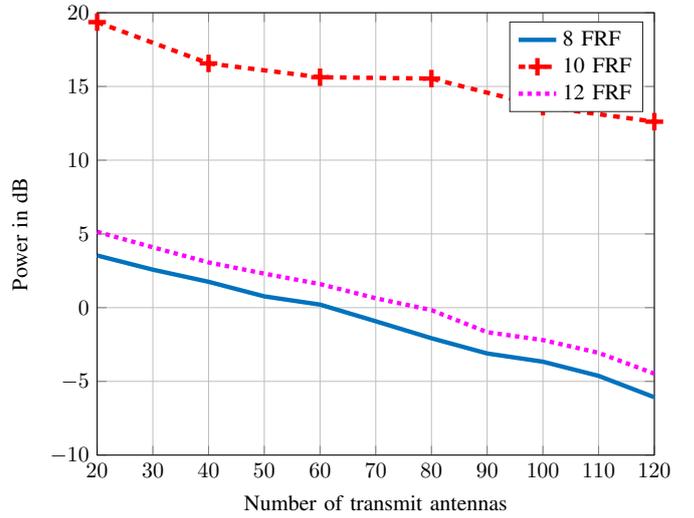

Fig. 6: Sum transmit power in dB for Rayleigh fading and $K = 10$.

TABLE II: Settings for the WINNER 2 Urban Macro Channel Model

| BS height | 25 m |
| --- | --- |
| Antenna array | ULA |
| Antenna distance | $\lambda/2$ |
| UE distribution | Uniform in 200 m radius |
| Center frequency | 2,5 GHz |

We next use the more realistic WINNER 2 Urban Macro channel model. It is implemented with the Quadriga MATLAB package and the settings are displayed in Table II. In Figure 7a we show the gain of adding $M_2 < K$ FRF-chains to the low-complex array. Throughout the whole antenna range we gain 6 dB in MSE. Continuing with Figure 7b we display the transmit power used in this scenario. The curve labeled KS shows the power for an array with only CRFs, FRF is the partial power of FRF-chains when combined with CRFs and HRF is the combined power of the arrays. Observe that the power usage of the FRF-chain antennas decreases with the number of total transmit antennas. Lower transmit power means higher energy efficiency and reduced cost as well as a reduced size of the overall array. In Fig. 8a and Fig. 8b the same scenario is shown, just with an additional 2 bits for phase information.

The MSE is much lower than the previous case with roughly the same distance between KS only and the HRF scheme with 120 antennas. In addition the transmit power

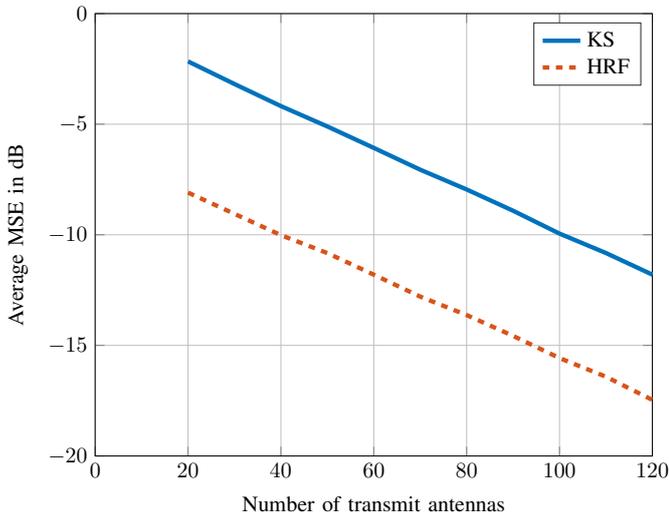
(a) Average MSE per UE for the WINNER channel, $K = 10$ UEs, 8 FRFs and 0 bit phase information.

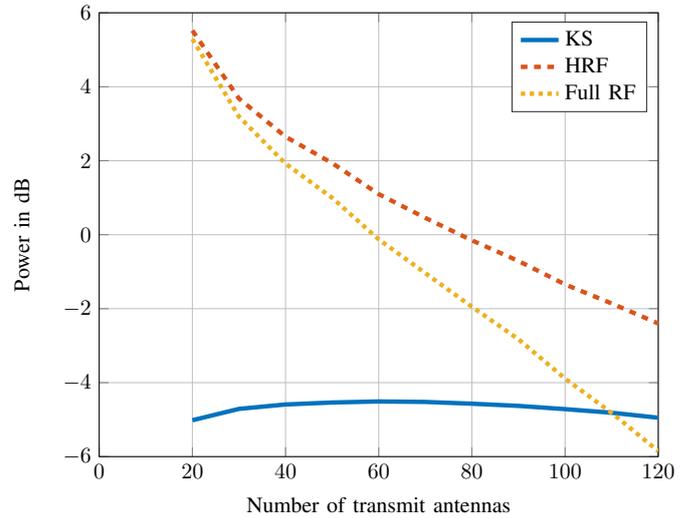
(b) Sum power for the WINNER channel, $K = 10$ UEs, 8 FRFs and 0 bit phase information.

Fig. 7: Performance for $m = 0$ phase bits.

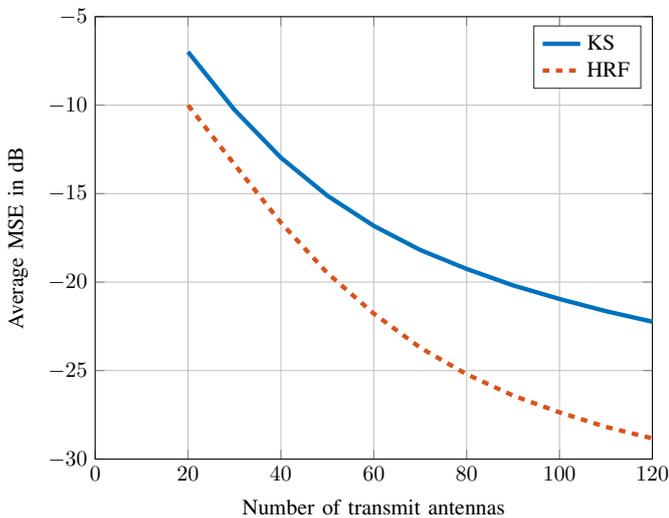
(a) Average MSE per UE for the WINNER channel, $K = 10$ UEs, 8 FRFs and 2 bit phase information.

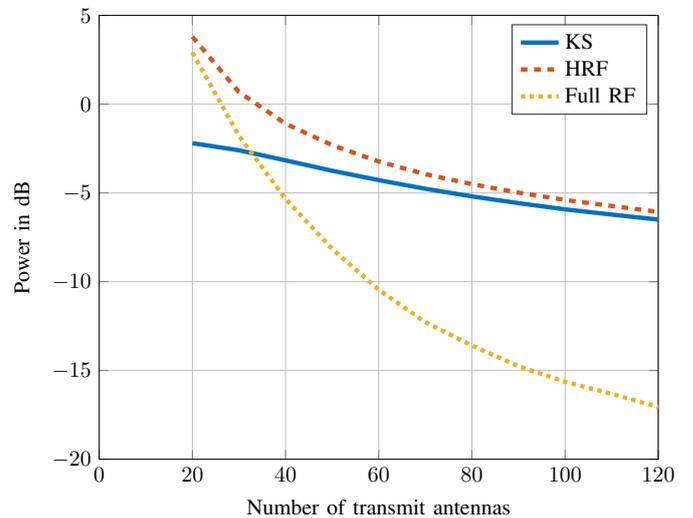
(b) Sum power for the WINNER channel, $K = 10$ UEs, 8 FRFs and 2 bit phase information.

Fig. 8: Performance for $m = 2$ phase bits.

of the FRF-chain is even lower and nearly negligible as compared to transmit power of the whole array. The cost of the RF-chain would increase only marginally due to the addition of phase information.

Finally we present a complexity analysis in Fig. 9. The upper bound shows the number of FLOPs needed if the KS algorithm runs through all antenna options. This will typically not happen as the algorithm at some point will not find new channel coefficients that decrease the error any further. The solid curves show the number of FLOPs needed for our simulations, where about one half of the available antennas are used. The number of FLOPs is calculated with a counting function from [16]. Considering an orthogonal frequency division multiple access (OFDMA) system like LTE for a bandwidth of 20 MHz and a sample rate of 33 Msample per second, the complexity of this algorithm needs further simplification for a low-cost implementation.

## IV. CONCLUSION

We have proposed a novel implementation of massive MIMO antenna arrays. Hybrid beamforming - the conventional alternative - might become either complex or face some performance limitations. An HRF solution that combines a large number of low-cost low-size CRF-chains with a limited number of FRF-chains maintains full MU MIMO scheduling flexibility as well as massive MIMO benefits like improved energy efficiency. We proposed a KS-like algorithm that achieves near optimal results with respect the MSE. The system can achieve signal to interference ratios supporting the highest LTE modulation and coding schemes. The combined scheme can improve the MSE even when the number of FRF-chains is smaller than the number of UEs. This is an interesting use case for booster antennas added to current macro sites, having only four to eight FRF-chains. Future work will focus on channel estimation, further reductions in processing overhead, optimum combining schemes and system level aspects.

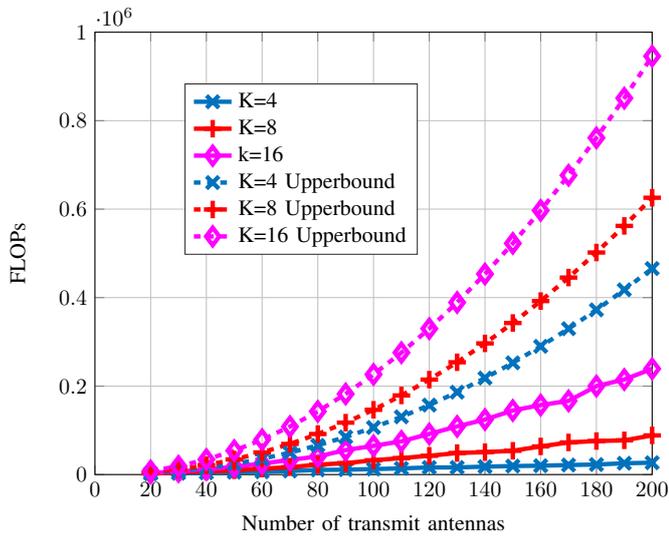

Fig. 9: FLOPs needed to find the solution for each channel realization and symbol without phase information.